\documentclass[12pt,preprint]{aastex}

\shorttitle{Extinction with 2MASS}
\shortauthors{Cambr\'esy et al.} 

\begin{document}

\title{Extinction with 2MASS: star counts and reddening toward the North
America and the Pelican Nebulae}

\author{L. Cambr\'esy}
\affil{Infrared Processing and Analysis Center, Jet Propulsion Laboratory,\\
California Institute of Technology, M/S 100-22, Pasadena, CA 91125}
\email{laurent@ipac.caltech.edu}

\author{C. A. Beichman}
\affil{Jet Propulsion Laboratory, California Institute of Technology, \\
M/S 180-703, Pasadena, CA 91109}
\email{chas@pop.jpl.nasa.gov}

\author{T.H. Jarrett and R. M. Cutri}
\affil{Infrared Processing and Analysis Center, California Institute of
Technology,\\ M/S 100-22, Pasadena, CA 91125}
\email{jarrett@ipac.caltech.edu,roc@ipac.caltech.edu}

\begin{abstract}
We propose a general method for mapping the extinction in dense molecular
clouds using 2MASS near-infrared data. The technique is based on the
simultaneous utilization of star counts and colors. These two techniques
provide independent estimations of the extinction and each method reacts
differently to foreground star contamination and to star clustering. 
We take advantage of both methods to build a large scale
extinction map ($2.5^\circ \times 2.5^\circ$) of the North America-Pelican
nebulae complex. With $K_s$ star counts and $H-K_s$ color analysis the
visual extinction is mapped up to 35~mag. Regions with visual extinction
greater than 20 mag account for less than 3\% of the total mass of the cloud.
Color is generally a better
estimator for the extinction than star counts. Nine star clusters are
identified in the area, seven of which were previously unknown.

\end{abstract}

\keywords{ISM: clouds --- dust, extinction --- ISM: individual (North America
Nebula, Pelican Nebula) --- infrared: ISM --- stars: formation}

\section{Introduction}
The knowledge of the extinction in molecular clouds is essential for a
wide range of purposes. The extinction traces the dust distribution and thus provides
an estimate of the column density of hydrogen that can be compared to the
far-infrared or submillimeter emission from dust and to molecular emission
lines. Fundamental properties are derived from these comparisons.
\citet{SAB+01} have found evidence of dust evolution in a translucent cloud
in Taurus using extinction from $J$-2MASS star counts and submillimeter data
from the balloon-borne experiment PRONAOS. 
\citet{KAL+99} have compared the C$^{17}$O line with extinction derived from $H-K$
color and have found a depletion of CO for $A_V>10$ which indicates the
interaction between dust an gas in high density regions.
The dust distribution also gives information about the fragmentation
processes which may be responsible for the shape of the initial mass function
\citep{PNJ97} through turbulent motions in the interstellar medium
\citep{PJN97}.
All these results were obtained with near-infrared data which are better
adapted than optical wavelengths for extinction studies.
2MASS provides accurate ($\sim 3$\%) $J$, $H$ and $K_s$ photometry for the
whole sky.
The $K_s$ band is about 10 times less absorbed than the $V$
band making possible to analyze the dust distribution for visual
extinctions as large as 40~mag. Since the dust both absorbs and reddens the
starlight, the analysis of the star density and star color variations give
two independent estimates of the extinction.
In this paper, we take advantage of these two techniques 
to measure the extinction and understand better the limitations of each
method.

The first section of this paper describes the methods
for mapping the dust distribution based on $K_s$ star counts and $H-K_s$
reddening. The comparison of the two maps leads to the identification of
foreground stars and stellar clusters.
In a second section, we apply our technique to the North America-Pelican
complex which is located in the galactic plane at an intermediate distance
($\sim 580$~pc). By analyzing the statistical and systematic uncertainties in
the star count and in the color extinction maps we merge these two maps into
a more reliable representation of the cloud. We also propose the
identification of new stellar clusters in this complex field.
In the appendix, we discuss the comparison between 2MASS data and a synthetic
model for stellar populations.

\section{Extinction mapping}
\subsection{Star counts}
\label{counts}
Variation of the star density across the sky was the first observable
indication of extinction, i.e. of presence of dust in the Galaxy. Star
counts are therefore the oldest method of mapping the interstellar cloud
extinction \citep{Wol23}. The method consists of counting stars in magnitude
intervals in each cell of a regular grid in an obscured field and comparing
this number with the counts in a reference field, supposedly free of
absorption. To improve the spatial resolution, \citet{Bok56}
made counts up to the limiting magnitude rather than in
magnitude intervals. Manual techniques were used on Schmidt plates for many
decades \citep{Dic78,CB84,Mat86,GSL88,AV96}.
More recently \citet{CEC+97} proposed a variant of the classical method by
replacing the regular grid by an adaptive grid. They used infrared data from
the DENIS survey \citep{Epc97} and an automatic counting technique that is
possible with digital data. The adaptive grid optimizes the angular resolution
with respect to the local star density. This is useful when there are
significant fluctuations of density due to large extinction gradients or the
presence of star clusters.

In its simplest form, star count method assumes that all stars are behind the
cloud and that the stellar population is homogeneous for the whole field.
Moreover, the star density is supposed to be uniform within a cell. Cloud
structures on angular scales smaller than the resolution of unit cell can
lead to underestimates of the extinction \citep{Ros80,TBD97}.
From star counts in the $K_s$ band, the visual ($V$) extinction is derived
from star density ($D$) as follows:
\begin{eqnarray}
A_{K_s} &=& \frac{1}{a} \log \frac{D_{\rm ref}}{D} =
	\mathcal{Z} + \frac{1}{a} \log \frac{1}{D} \label{cnt1}\\
A_V     &=& \frac{A_V}{A_{K_s}} \times A_{K_s}
\end{eqnarray}
where $a$ is the slope of the $K_s$ luminosity function and $D_{\rm ref}$ the
density in a reference field. Since the reference density depends on the
position (mainly on the galactic latitude), we use a {\em zero point function},
$\mathcal{Z}$, which can be approximated by a constant value only for
small areas (typically less than 1 deg$^2$).
For larger areas, $\mathcal{Z}$ is estimated using a fit of the uncalibrated
$A_{V}$ versus galactic position for unobscured regions close to the cloud.

\subsection{Reddening}
\label{color}
Dust that obscures also 
reddens starlight. Measuring the reddening of stars behind a cloud gives the
extinction along the line of sight. The spatial resolution of the derived
extinction is very high and corresponds to the apparent size of the
stellar disk, but the measurements are strongly undersampled. The reddening
is the difference between the observed star color and the intrinsic color
corresponding to the star spectral type.
\citet{FLW82} used $J-K$ color to map the extinction toward fields in $\rho$
Ophiuchus and Taurus. \citet{JHB84} used the same color to map the extinction
in a Bok globule toward the Southern Coalsack.
Until recently, the necessity of knowing the spectral type precluded a wide
use of this method. However, with the advent of large digital surveys like
2MASS, statistical studies become possible where the knowledge of the spectral
type is replaced by the measure of the mean color in an unobscured field.
Using 2MASS prototype data, \citet{BJ94} derived the extinction toward
the Taurus cloud by comparing $J-H$ and $H-K_s$ colors to a galactic model of
stellar populations. \citet{LLCB94} proposed a method
based on the $H-K$ averaged color in a regular grid to map highly extinguished
clouds with visual extinction up to 30-40 mag \citep{ALL+98,LAL99}. Using DENIS
data \citet{SGS+99} mapped the extinction toward the galactic center with a
sophisticated method based on the isochrone position in a $J-K_s / K_s$
color-magnitude diagram.

As for star counts, a homogeneous population is assumed for the reddening
analysis. Stars must lie behind the cloud and the reddening is supposed to be
uniform within a cell. For example, the study of \citet{SGS+99} in the
direction of the galactic center was restricted to RGB/AGB stars; they
selected a specific stellar population at a large distance (behind
the galactic center).

We have chosen to use the $H-K_s$ color because it presents a small
intrinsic dispersion over typical stellar spectral types. Typically, we have
$0<H-K_s<0.37$ \citep{BB88} and a single peak in the color distribution.
The $J-K_s$ color has separate peaks for dwarf and giant stars which makes it
difficult to convert a mean color to an extinction.

We propose here an adaptive variant of the Lada's method. The size of our
cells change with the local stellar density and is varied to contain
always the same number of stars detected in both $H$ and $K_s$ 2MASS bands.
The motivation for using adaptive cells is the same as that mentioned for
star counts: to optimize the angular resolution for large variations of
extinction and increase the resolution around stellar clusters.
Visual extinction is derived as follows:
\begin{eqnarray}
E_{H-K_s} &=& (H-K_s)_{\rm obs} - (H-K_s)_{\rm int} \\
A_V  &=& \left(\frac{A_H}{A_V} -
		\frac{A_{K_s}}{A_V}\right)^{-1} \times E_{H-K_s}\label{eq4}
\end{eqnarray}
where $(H-K_s)_{\rm obs}$ is the observed median color in the cell and
$(H-K_s)_{\rm int}$ is the observed median color in the reference field and
represents the intrinsic median color for unreddened stars. 
This differs from the Lada's method in which mean color is used rather
than median color. Our choice is driven by the foreground star correction
that is discussed in section \ref{effect_foreground}.
Color is less sensitive to galactic latitude than star counts and a
constant characteristic value (e.g. zero point of extinction)
is correct for several square degrees. On larger scales, the
dependence with the galactic latitude (and longitude) must be taken into
account with a zero point which is a function of direction.

\subsection{Simultaneous use of the two methods}
\subsubsection{Calibration}
To use both star counts and reddening to study a cloud it is necessary to
ensure that the calibrations are consistent. First, the spatial resolution
must be identical, i.e. same cells must be used. Practically, the cell size
is adapted to contain a number of stars, $N$, detected in both $H$ and $K_s$.
The $K_s$ star density is measured in the cell (number of $K_s$ stars divided by
the cell surface area) and the $H-K_s$ color of the $N$ stars is represented by
their median value.

Star counts and colors are converted into extinction using 
equations~(\ref{cnt1})-(\ref{eq4}). Since the zero point of the color-based
extinction map is more reliable than that of the star counts-based map
due to a weaker dependence on galactic latitude, the star counts map
is adjusted to the zero point of color map. The larger the area free from
absorption used, the better is the adjustment.\\
For a small field ($\sim 1$~deg$^2$), the difference between the maps is simply
an offset. For a large cloud ($\lesssim 100$~deg$^2$), a linear fit of the
difference between the two maps versus the galactic latitude is required.
For still larger sizes, variation with the galactic longitude may also occur
and the galactic latitude fit may no longer be linear.

\subsubsection{Expected differences}
\label{expect_diff}
Once the extinction maps are constructed with a consistent calibration, 
significant differences are still expected because each method has its
own behavior when the {\it a priori} assumptions are not valid.

\subsubsubsection{Effect of foreground stars\label{effect_foreground}}
\paragraph{Star counts.}
Foreground stars are not affected by the cloud extinction and their number
density, $N^f$, does not vary; whereas the background star density, $N^b$, is
affected by the extinction as described in equation~(\ref{cnt1}): $N^b = N_0^b
\times 10^{-a\, A_{K_s}^{\rm real}}$, where $a$ is the $K_s$ luminosity
function slope and the $0$ subscript refers to zero-extinction stellar density. 
The observed extinction can be written:
\begin{eqnarray}
A_{K_s}^{\rm obs} &=& \frac{1}{a} \log \frac{N_0^b + N_0^f}{N^b + N_0^f} \\
		  &=& \frac{1}{a} \log \frac{N_0^b + N_0^f}
		  	{N_0^b \times 10^{-a\, A_{K_s}^{\rm real}} + N_0^f}
\end{eqnarray}
Let $X$ be the fraction of background stars, $X=N_0^b / (N_0^b + N_0^f)$,
we have:
\begin{eqnarray}
A_{K_s}^{\rm obs} &=& -\frac{1}{a} \log \left(X\, 10^{-a\, A_{K_s}^{\rm real}}
			+ (1-X)\right)\label{eq7}
\end{eqnarray}
Foreground stars reduce the density contrast between low and high
extinction regions. For a given $X$, the measured extinction cannot exceed 
$-1/a \times \log (1-X)$, even for an infinite real extinction. For example,
with $X=0.9$ the observed visual extinction cannot exceed 26~mag from
$K_s$ star counts (10.5~mag from $J$ star counts and 3~mag from $V$ star
counts). 

\paragraph{Reddening.}
Since the color in a cell is represented by the median color of the stars it
contains, the contamination by foreground stars makes the derived extinction
drop to zero when more than half of the stars are in front of the cloud
(i.e. when the median color corresponds to the foreground star color).
This happens at the wavelength $\lambda$ for
$A_\lambda = -1/a \log [(1-X)/X]$. In the $H$ band, which is the
most affected by extinction, the turn over occurs at $A_V \approx 16$~mag for
$X=0.9$.

\begin{figure}
\epsscale{0.5}
\plotone{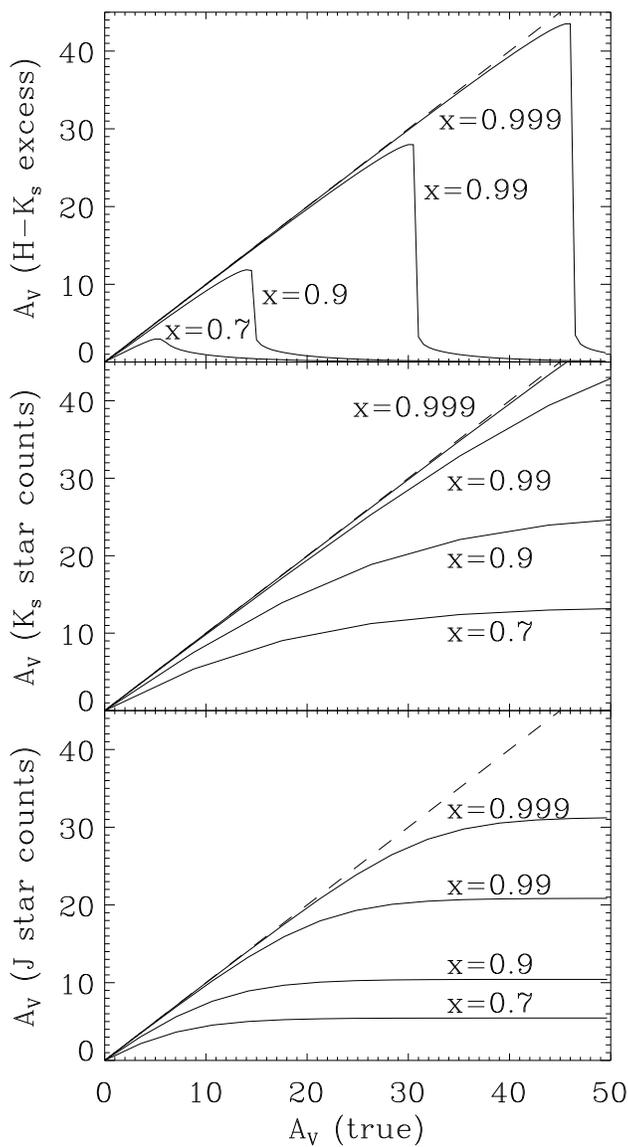}
\caption{Effect of the foreground star contamination on the {\em measured}
	extinction versus the {\em true} extinction for different degree of
	contamination ($X= N^{\rm background} / N^{\rm total}$) and
	different methods ($H-K_s$ color, $K_s$ star counts and $J$
	star counts).}
\label{foreground_colcnt}
\epsscale{1.0}
\end{figure}

\paragraph{Foreground star suppression.}
In Figure~\ref{foreground_colcnt}, the {\em observed} extinction is plotted
versus the {\em true} extinction for a fraction of background stars $X$.
The star count plots are obtained using equation~(\ref{eq7}). The color
excess plot is obtained assuming a color distribution characterized by 
$H-K_s = 0.17 \pm 0.1$~mag that corresponds to the value obtained with
the stellar population model described in the appendix, for a direction
close to the galactic plane ($l=85^\circ$, $b=-1^\circ$).
As expected, $K_s$ star counts give better results for high extinction than
$J$ star counts because of the relative sensitivity to the extinction
($A_J / A_{K_s} \approx 2.5$).
The color excess technique is definitely more robust to foreground star
contamination than star counts until the number of background stars drops to
equal the number of foreground stars.

Consequently, for high extinction, $A_V \sim 20$~mag and $X=0.9$, $K_s$ star
counts give a much higher extinction ($\approx 15$~mag) than does $H-K_s$
reddening (almost 0~mag). The star count extinction map would show an
extinction peak where the reddening map would show a hole in the cloud.
Using the two maps allows one to show that the extinction is so high that
there are more foreground stars than background stars in a particular direction.
The foreground star density can be estimated from these directions using
the $J-K_s$ color, which is the most sensitive to extinction. Once the
foreground star density is known,
we can remove these objects for each direction assuming that they are the bluer
(i.e. lower $J-K_s$) population. This hypothesis is verified only for high
extinction regions, where the reddening is larger than the total uncertainties
in the two maps. For low extinction, removing the bluer stars introduces a bias
that will be discussed later (\S~\ref{result}).
When foreground stars are removed from the catalog, the star counts and the
reddening analysis must be iterated to build reliable extinction maps
for which the foreground star contamination is corrected.

\subsubsubsection{Young stellar clusters}
\label{yso}
Molecular clouds are star forming regions and young embedded star clusters are
to be expected in such areas. 
A cluster produces an artifact in the star count extinction map:
the excess density is interpreted as a hole in a cloud. On the other hand,
such clusters generally contain stars with an intrinsic infrared excess due
to an envelope or an accretion disk (especially for low-mass stars) which
leads to an extinction peak in a $H-K_s$ color extinction map. 

This is the opposite situation that was encountered for the contamination by
foreground stars: the extinction appears larger in the color map than in the
star count map. Directions for which the star density is significantly too
large compared to the color suggest the presence of star clusters. A similar
approach is developed in \citet{Car00} who identifies young clusters by
comparing 2MASS $K_s$ star counts to stellar densities predicted by a
model toward Perseus, Orion and Monoceros R2 molecular clouds and using CO
line to correct for the extinction.

\section{Application to the North America-Pelican nebulae}

\subsection{2MASS data}
The 2MASS survey provides near--infrared photometry in the $J$, $H$ and $K_s$
filters. At the time of this paper data covering, 48\% of the sky has been
released through the 2$^{\rm nd}$ Incremental Data Release \citep{CSV+00}
and 100\% of the sky has been observed. Since data analyzed here are not in
the 2$^{\rm nd}$ release, we have extracted the catalog directly from the
point source working database which contains sources extracted in each Tile
(a 2MASS unit of $8.5'\times 6^\circ$ obtained by the co-addition of 274
images).
As distinguished from the 2$^{\rm nd}$ release Point Source Catalogue, the
working database contains multiple entries for the same star in Tile overlap
regions, low signal-to-noise sources, and artifacts flagged during the
processing. We have extracted from the point source working database sources
that satisfy the following criteria which are similar to the criteria used to
generate the point source catalogue:
\begin{enumerate}
\item photometric uncertainty $\sigma<0.25$
\item signal-to-noise greater than 7 in at least one band
\item no filter glint flag (ghost sources that occur at
specific geometric locations relative to bright stars)
\item no diffraction spike flag
\item persistence probability due an echo of a previously observed bright
star $<0.5$ in the 3 bands simultaneously
\end{enumerate}

In this filtering, we kept sources bright enough to be accepted as a real
source but weakly contaminated by a diffraction spike, and sources with high
PSF fitting $\chi^2$ (generally due to confusion between two stars). We give
preference to completeness rather than reliability because our analysis is
statistical and involves star counts.

Redundant source entries in the Tile overlaps were removed by filtering out
sources within $3''$ of bright stars (brighter than 8.5 mag in one band) and
within $1.5''$ of fainter stars. The final catalog contains about
$1.1 \times 10^6$ objects.

\begin{figure}
\epsscale{0.5}
\plotone{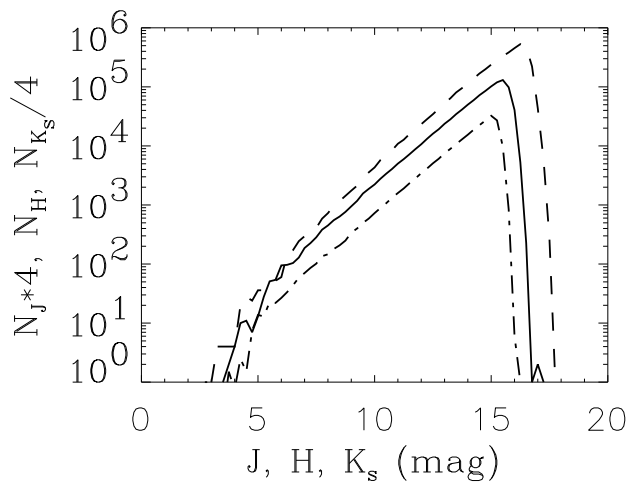}
\caption{Histogram distribution of the 2MASS magnitudes in the NAN
($J$: dashed line, $H$: solid line and $K_s$: dashed-dotted
line). Slopes of the luminosity functions are 0.36, 0.34 and 0.34 for $J$, $H$
and $K_s$, respectively.}
\label{completude}
\epsscale{1.0}
\end{figure}

Based on an estimate of completeness from the magnitude distribution,
Figure~\ref{completude}, we kept only stars brighter than $K_s<15.0$~mag for
the star count analysis and stars brighter than $H<15.5$~mag and $K_s<15.0$~mag
for the color analysis.
Analysis of repeated observations of calibration fields observed $\sim
100$ times indicates that the integral 2MASS completeness for these magnitudes
is $\sim 99$\% \citep{CSV+00}.
The density and color distribution of the selected stars does not show any
residual persistence of the 2MASS observing Tiles. This validates the
{\em catalog extraction} from the database and the adopted limiting magnitudes.
An incorrect treatment of multiple entries in the working database would have
led to the detection of the Tile overlaps in the density map while an
overestimation of the limiting magnitudes would have introduced a star density
level difference between Tiles, especially for those observed at different
dates, under different airmass and/or atmospheric conditions.
To derive extinction from $K_s$ star counts using equation~(\ref{cnt1}) 
the slope of the luminosity function, $a$, is needed. Its estimate from 
Figure~\ref{completude} is $a=0.34$.

\subsection{Cloud properties}
The cloud studied in this paper lies between two H{\small II} regions, the
North America nebula (NGC 7000) and the Pelican nebula (IC 5070). This
complex is also named W80 \citep{Wes58} and the dark cloud itself is known as
L935 \citep{Lyn62}. In the following, we use NAN to refer to the whole
complex. Visual extinction has been deduced from H$\alpha$ surface brightness by
\citet{GW79} who found a {\em maximum} visual extinction of about 8 mag. 
\citet{BS80b} performed $^{12}$CO observations of the cloud and proposed an
evolutionary scenario for the molecular cloud. They concluded that the cloud is
highly fragmented and contains an expanding molecular shell. \citet{WBB83}
proposed a model to described the whole complex based on radio continuum
observations. They gave a distance estimate of 500 pc. This distance has been
confirmed by \citet{SKVC93} who obtained $d=580$~pc using photoelectric
photometry of 564 stars in the direction of the complex.
The total mass of the cloud has been estimated from molecular observations
($^{12}$CO, $^{13}$CO and H$_2$CO) to be $5 \times 10^4$ $M_\sun$ \citep{FW93}.

\begin{figure}
\plottwo{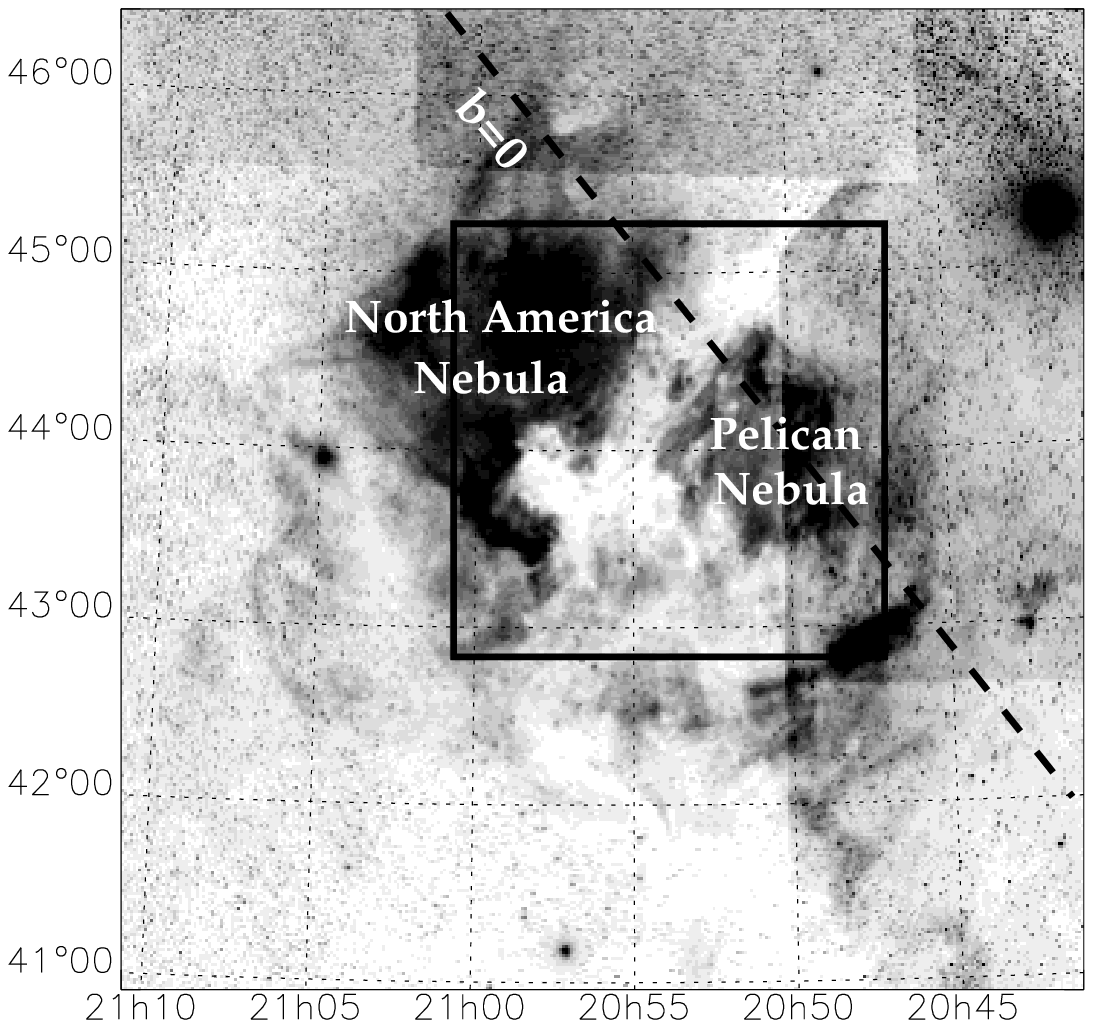}{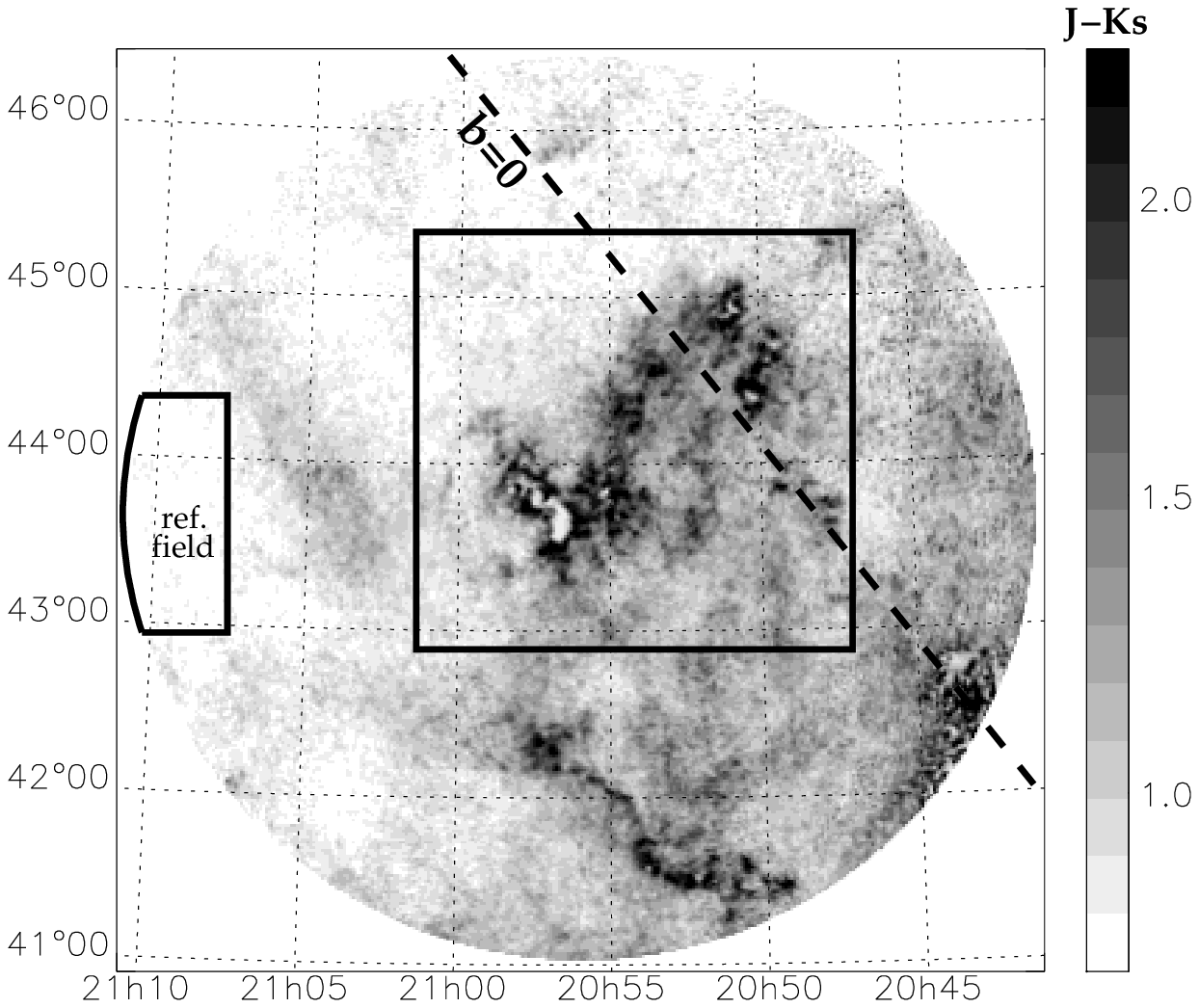}
\caption{Digitized Sky Survey optical image (left) and 2MASS $J-K_s$ map
	(right) of the NAN. The square shows the area studied in the paper.
	The reference field is used to set the zero point of extinction.
	Coordinates are expressed in the equatorial J2000 system.}
\label{mapJK}
\end{figure}

\citet{Her58} discovered 68 stars having the H$\alpha$ line toward the
complex, suggesting the existence of a population of Be (intermediate mass)
and T Tauri (low mass) stars. \citet{DCA84} reobserved these sources to
provide more accurate astrometry. \citet{Wel73} and \citet{Mar80} discovered
additional H$\alpha$ emission stars in the NAN.
The total number of previously known young stars in the square area defined
in the Figure~\ref{mapJK} is 119.

To clearly delimit the dark cloud, we performed the star counts and
the color analysis on a large area around the cloud: $5.6^\circ$ diameter
centered on ${\rm \alpha_{J2000}=10^h56^m00^s}$, 
${\rm\delta_{J2000}=44^\circ43^m43^s}$.
However, the extinction study is restricted to a $2.5^\circ \times 2.5^\circ$
area centered on ${\rm \alpha_{J2000}=20^h54^m05^s}$, 
${\rm \delta_{J2000}=44^\circ13^m57^s}$ ($l=84.9$, $b=-0.41$).
The optical image and $J-K_s$ map of the large region are presented in
Figure~\ref{mapJK}. $J-K_s$ color is more sensitive to low extinction than
$H-K_s$ and shows the general shape of the dark cloud. The reference field to
set the zero point of extinction has been chosen in the area of lower $J-K_s$
value. It contains $20,000$ stars detected in both $H$ and $K_s$.

\subsection{Results}
\label{result}
\begin{figure}
\epsscale{0.5}
\plotone{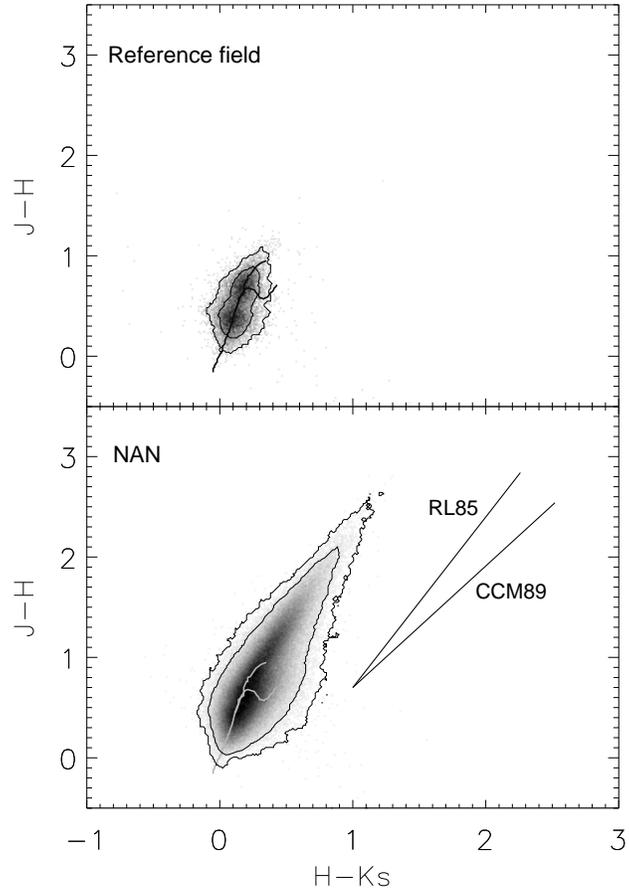}
\caption{Color color diagram for the reference field (top) and the NAN
(bottom) directions. RL85 and CCM89 are the
reddening vectors for \citet{RL85} and \citet{CCM89} extinction laws,
respectively. Dwarf and giant sequences are overlaid \citep{BB88}.}
\label{ccRL85_CCM89}
\epsscale{1}
\end{figure}

In the following, we assume the \citet{RL85} extinction law rather than the most
widely used \citet{CCM89} law because the former is in better agreement with the
reddening vector slope measured in the 2MASS color-color diagram ($H-K_s/J-H$)
obtained for the NAN (Figure~\ref{ccRL85_CCM89}). We have measured a slope of
$1.61\pm0.32$ and \citet{RL85} and \citet{CCM89} obtained 1.70 and 1.21,
respectively. We will therefore use the following values:
$A_J/A_V=0.282$, $A_H/A_V=0.175$ and $A_{K_s}/A_V=0.112$.

\subsubsection{Extinction}
\begin{figure}
\epsscale{0.5}
\plotone{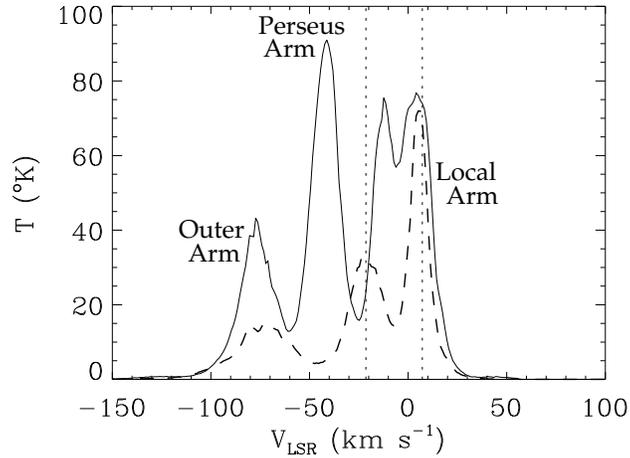}
\caption{Velocity components toward the NAN (solid line)
and the reference field (dashed line) from the Leiden/Dwingeloo survey. The
vertical dotted lines show the velocity range of the NAN
(in the Local Arm).}
\label{nanHI}
\epsscale{1}
\end{figure}

\subsubsubsection{Reddening method}
The extinction map of the NAN is obtained using
$H-K_s$ color with the method described in \S~\ref{color}. The average
color in the reference field provides the absolute calibration of the
extinction (zero point). The average color in the reference region is
$(\overline{H-K_s})_{\rm int} = 0.16$, with a dispersion ($1~\sigma$) of
0.12~mag which corresponds to $A_V=1.9$~mag. 
The dispersion represents the sum of the photometric uncertainties,
the intrinsic dispersion due to the average over all spectral types 
and a possible contamination by a residual absorption is the field.
This dispersion is consistent with the one obtained using the galactic
model presented in the appendix (0.1~mag), suggesting that there is no
significant extinction contamination in the reference field.
To estimate the systematic error resulting from a possible residual absorption
one can use the hydrogen column density.
Molecular hydrogen column density can be derived from the CO emission. For
this field $W({\rm CO})=0.03$~K~km~s$^{-1}$ \citep{DHT01} which corresponds
to $N({\rm H}_2)= 5.4 \times 10^{18}$~cm$^{-2}$.
The atomic hydrogen column density is obtained from the 21~cm emission line 
and is $N(\mbox{H\sc i}) = 3.7\times 10^{21}$~cm$^{-2}$ 
\citep[Leiden/Dwingeloo survey,][]{BH94}.
This value represents the total hydrogen column integrated along the line of
sight.  A velocity analysis of the neutral gas allows one to separate the
different galactic components on the light of sight (Figure~\ref{nanHI}). 
Following the \citet{FGO83} results based on a H$\alpha$ study of the NAN,
the LSR velocity of the nebula is
$-7.1\pm5.5$~km~s$^{-1}$ with a FWHM of $28.6 \pm 0.6$~km~s$^{-1}$.
The fraction of H{\sc i} emission coming from this velocity range represents
$40\pm 7$\% of the total. This value is an upper limit since not all the gas in
this velocity range is associated with the complex.
For $[N(\mbox{H\sc i}) + N({\rm H}_2)]/A_V = 1.87\times 10^{21}$ \citep{SM79}
the upper limit for the visual extinction in the reference field is about
$0.78\pm0.14$~mag (only 0.4\% is associated with the molecular gas).

\subsubsubsection{Star count method}
The extinction is derived from $K_s$ counts as described in 
\S~\ref{counts}. Once the color map has been calibrated (previous
paragraph), the star count extinction map is adjusted using a linear relation
for the zero point versus the galactic latitude and longitude. 
For the NAN area, the stellar density is found to increase by $\sim 25$\%
per degree toward the galactic plane.
The cell size is allowed to vary to achieve the desired number of stars, 10 in
this work. Since the $H-K_s$ intrinsic color dispersion is low, this small
number gives good results for the color map
($\sigma(A_V) = 1.9/\sqrt{10} = 0.6$~mag).
However, this small number introduces a significant Poisson noise in the star
count map in which the standard deviation of the extinction is $\sigma=3.5$~mag.
For $A_V \gtrsim 20$~mag, the uncertainty is actually reduced for $K_s$ star
counts because the cell size is constrained by the $H$ band which is more
sensitive to extinction. Cells in high extinction regions contain 10 stars
simultaneously detected in $H$ and $K_s$ and more than 10 additional stars
detected only in $K_s$. The statistical uncertainty in the $K_s$ star count
map is reduced to $\sigma \approx 2.5$~mag when $A_V \gtrsim 20$~mag.

\begin{figure}
\epsscale{0.5}
\plotone{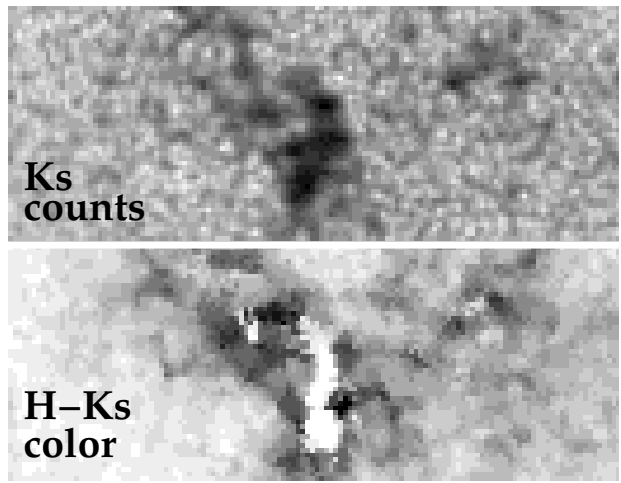}
\caption{Extinction maps from $K_s$ star counts (top) and $H-K_s$ color
	(bottom) for a $64' \times 24'$ field centered on the most obscured
	region of the dark cloud (${\rm \alpha_{J2000}=20^h56^m51^s}$,
	${\rm \delta_{J2000}=43^\circ42^m33^s}$). Black is for high extinction
	and white for low extinction. Extinction from star counts
	reaches 30~mag whereas the extinction from reddening at the same
	position is 0~mag due to the foreground star contamination (see
	\S~\ref{expect_diff}).}
\label{core}
\epsscale{1}
\end{figure}

\subsubsubsection{Foreground star density estimation}
Figure~\ref{core} shows the extinction from star counts and from reddening
in the most obscured portion of the field. The effect of the foreground star
contamination is obvious: the extinction derived from star counts reaches 
30~mag where the extinction derived from reddening drops suddenly to 0~mag.
In areas where $A_V({\rm star counts}) - A_V({\rm color}) > 20$
(i.e. $5.5\sigma$) we find 57 stars. Among them, we identify 15 stars which do
not show any $J-K_s$ excess (the expected excess when $A_V=20$ is
$E_{J-K_s}=3.4$).
The normalized density for this foreground star population corresponds to
$1900\pm500$ stars~deg$^{-2}$ and represents $\sim 6$\% of the total star
number for an unobscured field.

\begin{figure}
\epsscale{0.5}
\plotone{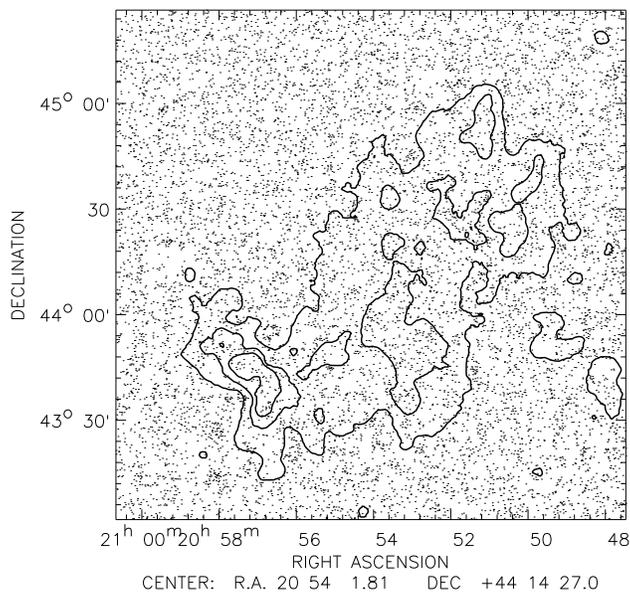}
\caption{Distribution of foreground stars with smoothed extinction contours
overlaid (contour values are 5, 10 and 15 $A_V$). Extinction is
derived from $H-K_s$ color.}
\label{fore_cont}
\epsscale{1}
\end{figure}

\begin{figure}
\plottwo{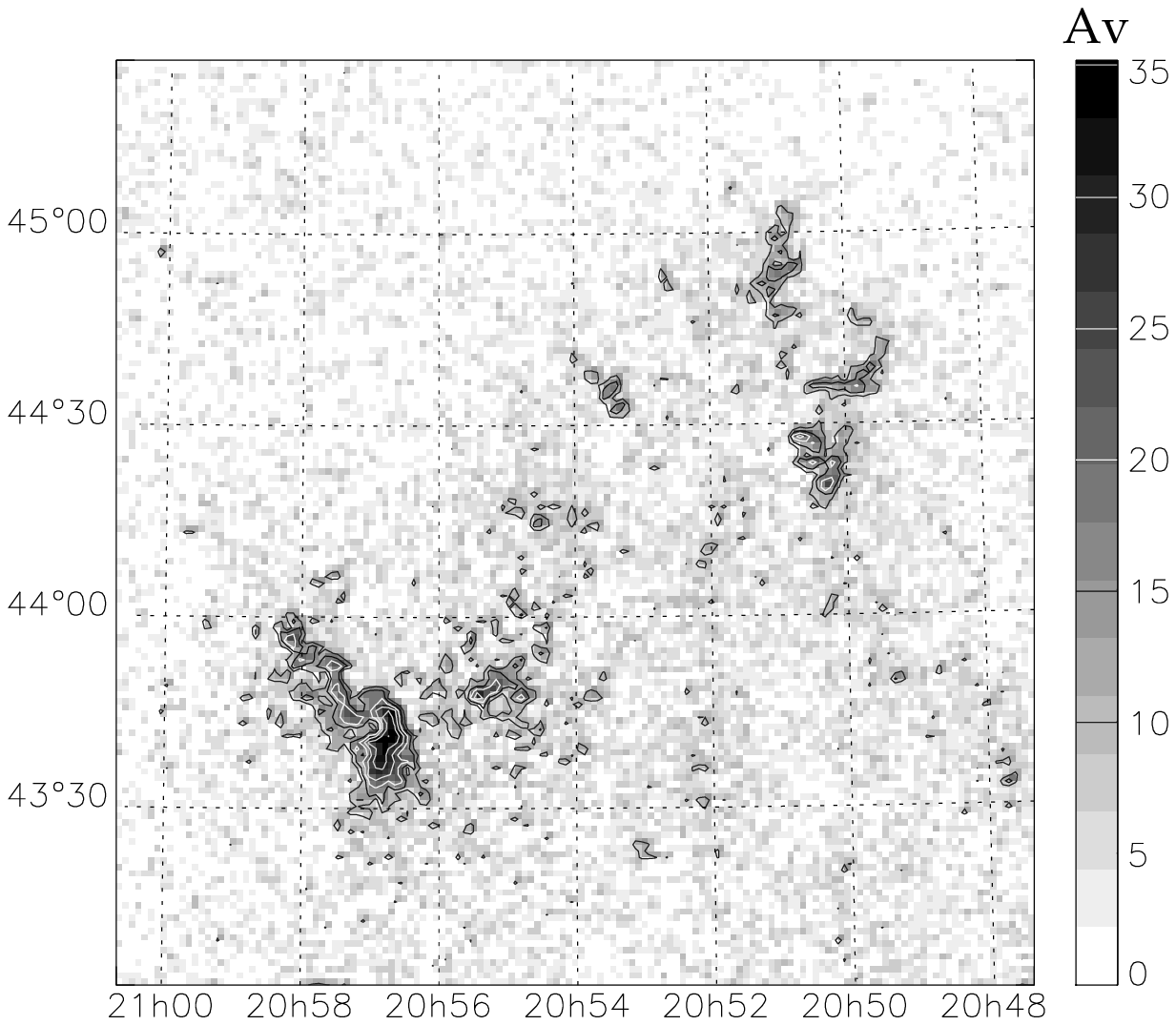}{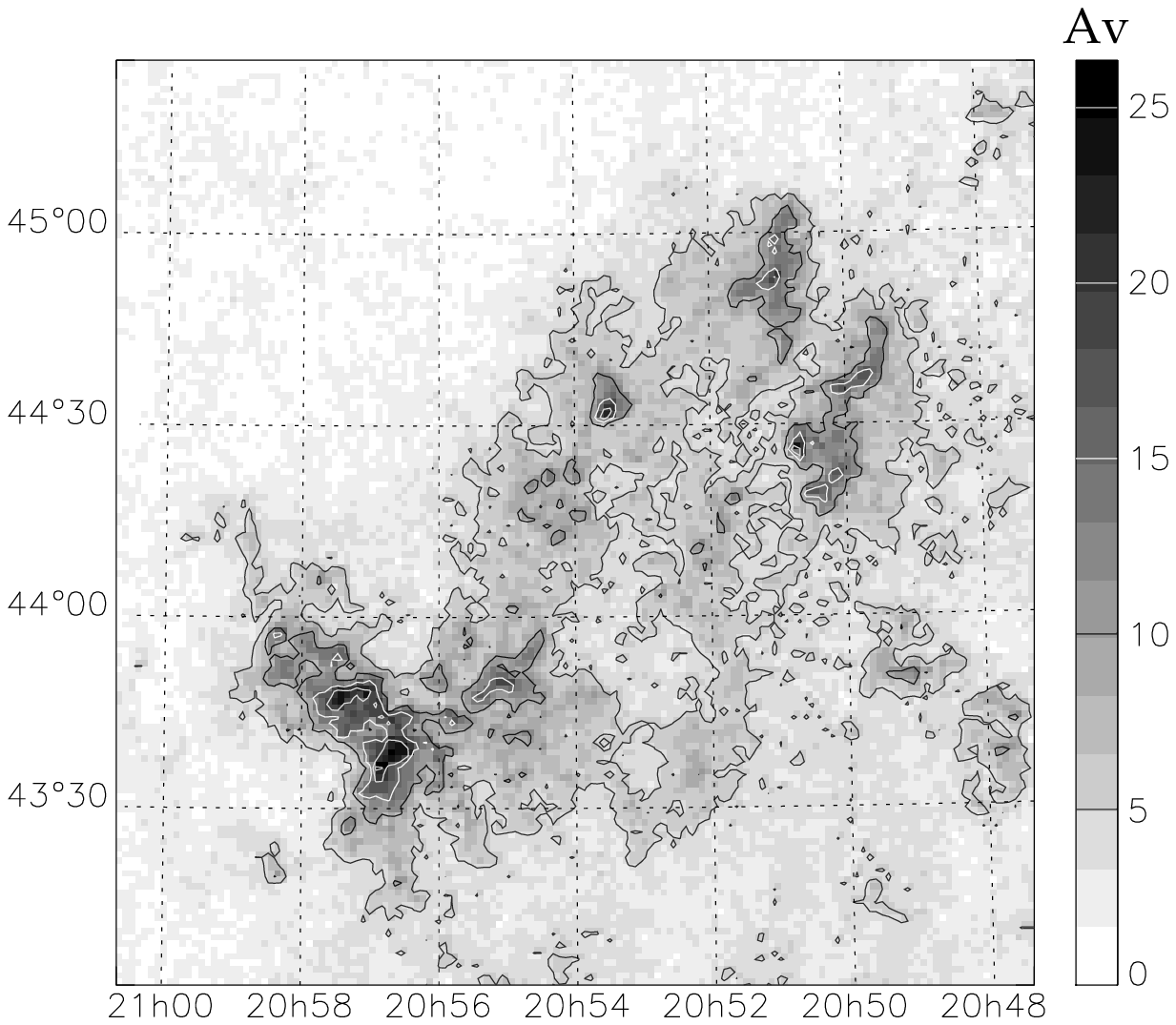}
\caption{Extinction maps derived from $K_s$ star counts (left) and from
	$H-K_s$ color (right) corrected for foreground stars. Coordinates
	are expressed in the equatorial J2000 system.}
\label{maps}
\end{figure}

\subsubsubsection{Foreground star correction}
Unfortunately, setting a color threshold to separate the two star populations
systematically yields a foreground star distribution that is strongly
correlated with the dust distribution. The degeneracy between background blue
stars and foreground red stars prevents the proper separation of the two
populations.
To correct for the foreground star contamination, we remove the 30 bluer
(i.e. lower $J-K_s$) stars per $7.5' \times 7.5'$ box in the whole field, 
corresponding to the foreground star density measured toward the densest part
of the cloud.
In high extinction regions this procedure removes foreground stars, but for
visual extinction lower than $\sim 15$~mag the degeneracy prevents a correct
separation of the two populations and a bias is introduced in the extinction
maps. 
The resulting foreground star distribution is presented in
Figure~\ref{fore_cont}. 
The iteration of the star counts and the reddening analysis with the
new version of the catalog provides the foreground star corrected extinction
maps (Figure~\ref{maps}).
The standard deviation of the difference between these maps and maps obtained
before the foreground star correction are 0.1 and 0.9~mag for the color
and the star count extinction map, respectively.
The main difference comes from the cloud core
which corresponds to a high extinction for both maps: visual extinction is
larger than 25~mag in the color map where it was 0~mag before the correction.

To estimate the error in the foreground star correction we compare
the distribution of these stars with the dust distribution.
The difference between the mean extinction in the whole field and the mean
value measured at the position of each supposedly foreground star should give
zero for uncorrelated distributions. We find $\Delta A_V=-0.26$~mag.
Simulations for an uniformly distributed star field yield a standard
deviation of $\sigma=0.03$.
The spatial distribution of the selected stars is still correlated with the
extinction and is therefore not consistent with a population of only
foreground stars.
The fraction of foreground stars in a $7.5' \times 7.5'$ box reaches 0.5 for
the darkest region and is less than 0.1 for $A_V \lesssim 15$. Consequently,
in regions for which the stellar color degeneracy exists, fewer than 10\% of
the stars are affected by the selection bias. For higher values, the
foreground star identification is straightforward (i.e. not statistical).
We conclude that the contamination by residual background stars in the sample
has negligible consequences on the extinction mapping compared to the
intrinsic dispersion of these maps:
the standard deviation between the corrected and the not-corrected
maps are 0.1 and 0.9~mag whereas the intrinsic uncertainty for the extinction 
maps are $\sigma=0.6$ and $\sigma=3.5$~mag for the color and the star count
map, respectively.

\subsubsubsection{Merging the two maps}
The uncertainty is lower from the color analysis than for star counts
for $A_V<15$. For higher values,
it is not clear whether star counts or color give the best answer.
The color method is more constraining because of the necessity to detect
stars in both $H$ and $K_s$ bands whereas only detections in $K_s$ are
required to build the star count map.
Since stars are detected only in the longer wavelength in the most extinguished
areas, more data are available for the star counts than for the color analysis.
Statistical uncertainties are lower in the star count map when the number
of $K_s$ stars is greater than 300 in a cell (that by construction always
contains 10 stars detected in both $H$ and $K_s$).
Using the extinction law, the mean color $\overline{H-K_s}=0.16\pm 0.01$
and the limiting magnitude difference for each band,
$H^{\rm lim}-K_s^{\rm lim} = 0.5$~mag, it would correspond to
$A_V \approx 70$~mag. {\em Therefore, the color analysis technique has a lower
statistical uncertainty everywhere in our field.}

The resolution of the extinction maps is adapted to the local star
density. It is $1.2'$ for $A_V=0$~mag, $1.9'$ for $A_V=10$~mag, $3.7'$
for $A_V=20$~mag and $7'$ for $A_V>30$~mag. Unfortunately, extinction is
probably no longer homogeneous for resolutions coarser than 0.5~pc ($3'$
for $d=580$~pc) and both methods underestimate the extinction in a non-linear
way. 
This effect can be simulated using an uniform spatial distribution for
the stars and a the luminosity functions observed with 2MASS toward the NAN
direction (see figure~\ref{completude}). One can then redden this simulated
stellar field with any known extinction profile, assuming all stars are
background and applying the 2MASS completeness cuts.
The simulation of a Gaussian peak of 30~mag of visual extinction with
a $50''$ FWHM observed at $100''$ resolution gives $A_V(H-K_s)=4.6$~mag
and $A_V(K_s\, {\rm counts})=8.0$~mag. A $90''$ Gaussian peak observed at a
$100''$ resolution gives $A_V(H-K_s)=16.5$~mag and
$A_V(K_S\, {\rm counts})=18.8$~mag.
When the extinction is not homogeneous over a cell, star counts lead to a
higher value than reddening. But both methods underestimate the peak value.
Thus, at the highest extinctions, using these techniques at the 2MASS
sensitivity underestimates the extinction. Deeper observations are needed
such as those presented in \citet{ALL01} with limiting magnitudes of about
20~mag in both $H$ and $K_s$.

\begin{figure}
\epsscale{0.65}
\plotone{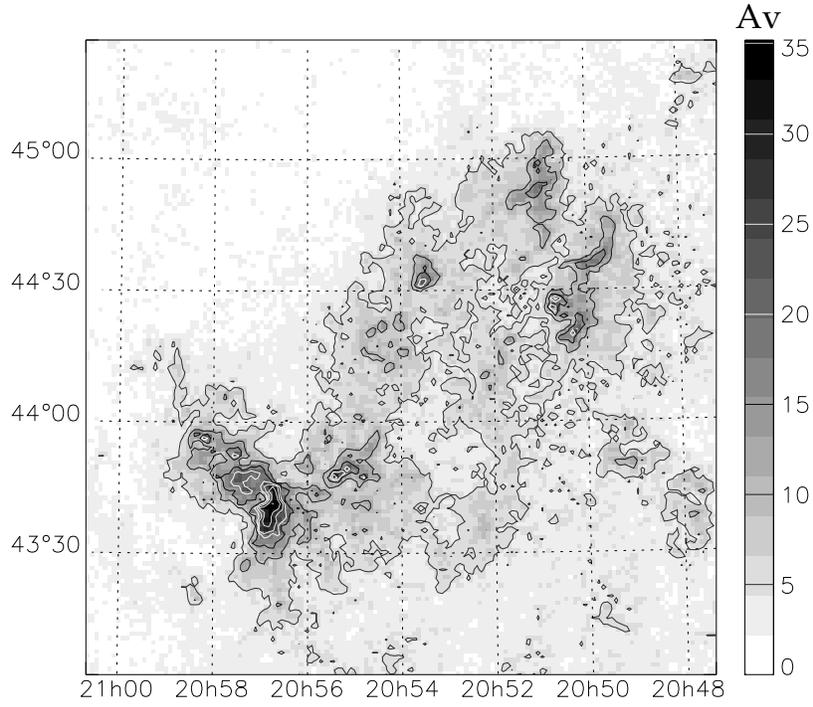}
\caption{Extinction map of the NAN obtained
by combination of the $H-K_s$ color map with the $K_s$ star count map.}
\label{finalmap}
\epsscale{1.0}
\end{figure}

Our analysis concludes that (1) the color method is more robust than stars
counts to the foreground star contamination. This contamination can be
corrected by
comparing color and count maps. (2) The star count extinction map has larger
statistical uncertainty than the color extinction map.
(3) The star count analysis is more robust than the color analysis to
extinction inhomogeneities within a cell, i.e. the systematic uncertainty is
smaller in the star count map for high extinction.
Because of these different characteristics, we decided to combine the two
extinction maps using reddening for $A_V<15$, star counts for $A_V>25$ and
a linear combination of the two maps for the intermediate extinction range:
\begin{equation}
A_V = x\times A_V(K_s\, {\rm counts}) + (1-x)\times A_V(H-K_s) \\
\end{equation}
where:
\begin{displaymath}
\begin{array}{lcr}
x=0 &\textrm{ for }& A_V(K_s\, {\rm counts})<15\\
x=\left[A_V(K_s\, {\rm counts})-15\right]/10 &
\textrm{ for }& 15\leq A_V(K_s\, {\rm counts})<25\\
x=1 &\textrm{ for }& A_V(K_s\, {\rm counts})\geq 25
\end{array}
\end{displaymath}
The final combined extinction map of the NAN is
presented in Figure~\ref{finalmap}. The maximum visual extinction is
$A_V=35$~mag whereas previous works at optical wavelengths estimate the
extinction to less than 10~mag \citep{GW79,SKVC93}. The 2MASS near--infrared
data allow deeper analysis of the dust distribution.

\begin{figure}
\epsscale{0.65}
\plotone{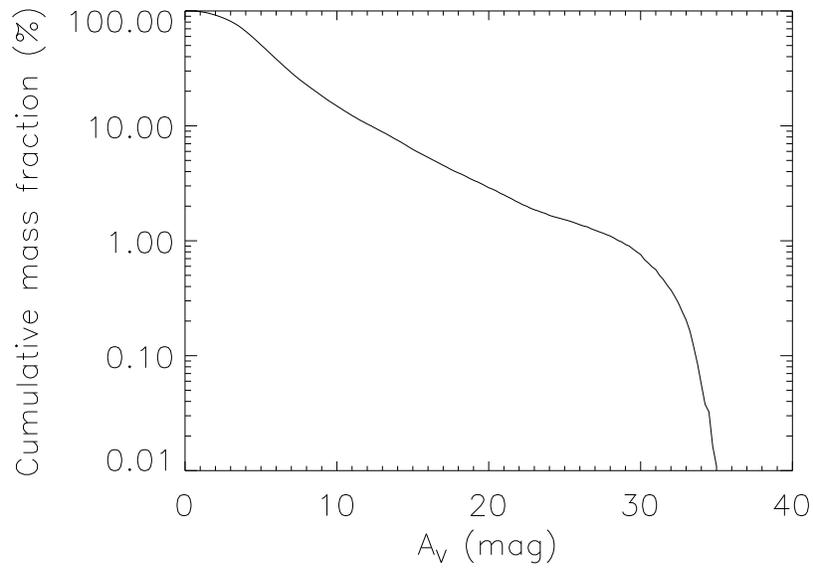}
\caption{Cumulative enclosed mass in iso-extinction contours versus visual extinction.}
\label{mass_spec}
\epsscale{1.0}
\end{figure}

\subsubsubsection{Cloud mass}
The cloud mass is derived from the extinction map using the following
relation \citep{Dic78}:
\begin{equation}
M = (\alpha d)^2 \, \mu \, \frac{N_H}{A_V} \sum_i A_V(i)
\end{equation}
where $\alpha$ is the angular size of a pixel map, $d$ the distance to the
cloud, $\mu$ the mean molecular weight corrected for the helium abundance,
and $i$ represents a pixel map. With the dust-to-gas proposed by
\citet{SM79}, $N_H/A_V=1.87\times 10^{21}$~cm$^{-2}$~mag$^{-1}$
($N_H=N_{HI}+N_{H_2}$) and a distance of 580~pc, we obtain a mass of 
$4.5 \times 10^4$~M$_\sun$. This value corresponds to about one fifth of the
Orion cloud mass \citep{MMMT86,Cam99a}. Uncertainties come from the distance
estimate for the cloud and the systematic error in the extinction. 
An uncertainty of a factor of two is believed to be a reasonable estimate. 
Figure~\ref{mass_spec} shows how the integrated mass distribution varies
with the extinction within the cloud. Regions with visual extinction greater
then 20 mag account for less than 3\% of the total mass of the cloud.
This suggests that the underestimation of the extinction in the cloud
cores has only a limited consequence on the total mass estimate.

\citet{BS80b} estimated the molecular cloud mass from CO observation to
3-$6 \times 10^4$~M$_\sun$ for a distance of 1~kpc. For 580~pc, it would 
correspond to 1-$2 \times 10^4$~M$_\sun$.
The discrepancy might come from the $N_H/A_V$ value or from the conversion
of CO density to $H_2$ density.

\subsubsection{Limitations of the method}
The cloud distance is the most critical parameter in the accurate application
of these methods. The farther away a cloud, the lower is the spatial
resolution and the larger the number of foreground stars.
Here, we have mapped visual extinctions of 15~mag at a resolution of $3'$
which corresponds to 0.5~pc ($d=580$~pc).
If the cloud were at 2~kpc, the $3'$ resolution would correspond
to 1.7~pc, equivalent to a study of nearby clouds such as
Taurus, Chamaeleon or $\rho$ Ophiuchus at a $40'$ angular resolution. 
This is the resolution of DIRBE/COBE and is probably close to the
{\em minimum acceptable} to derive properties within a cloud. As the
distance increases the number of foreground stars increases, as well,
further degrading extinction maps.

The 2MASS star density is about 4 times higher toward the galactic center
than for the NAN direction. Clouds toward such very high
density directions could therefore be mapped with a better resolution for a
same distance. In contrast, the star density is about 3.5 times lower toward
the anti-center and dramatically decreases for higher galactic latitudes
(density divided by 2 at $b\approx 5^\circ$). Thus, the maximum distance for
which molecular clouds can be mapped using 2MASS with this method cannot
exceed 3~kpc toward very high density directions such as toward the galactic
bulge.

\subsubsection{Star Formation}
\begin{figure}
\epsscale{0.5}
\plotone{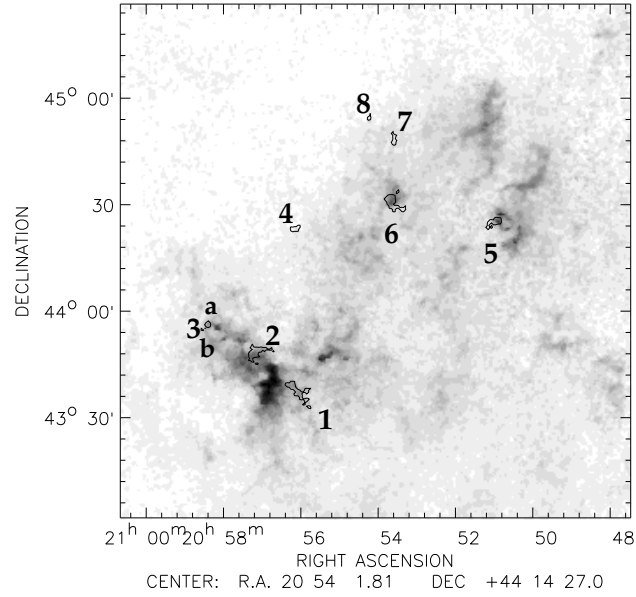}
\caption{Position of the nine clusters (contours correspond to
$A_V({\rm color})- A_V({\rm counts})>6$~mag) represented on the combined
extinction map of the cloud.}
\label{clust_pos}
\epsscale{1.0}
\end{figure}

\begin{figure}
\plotone{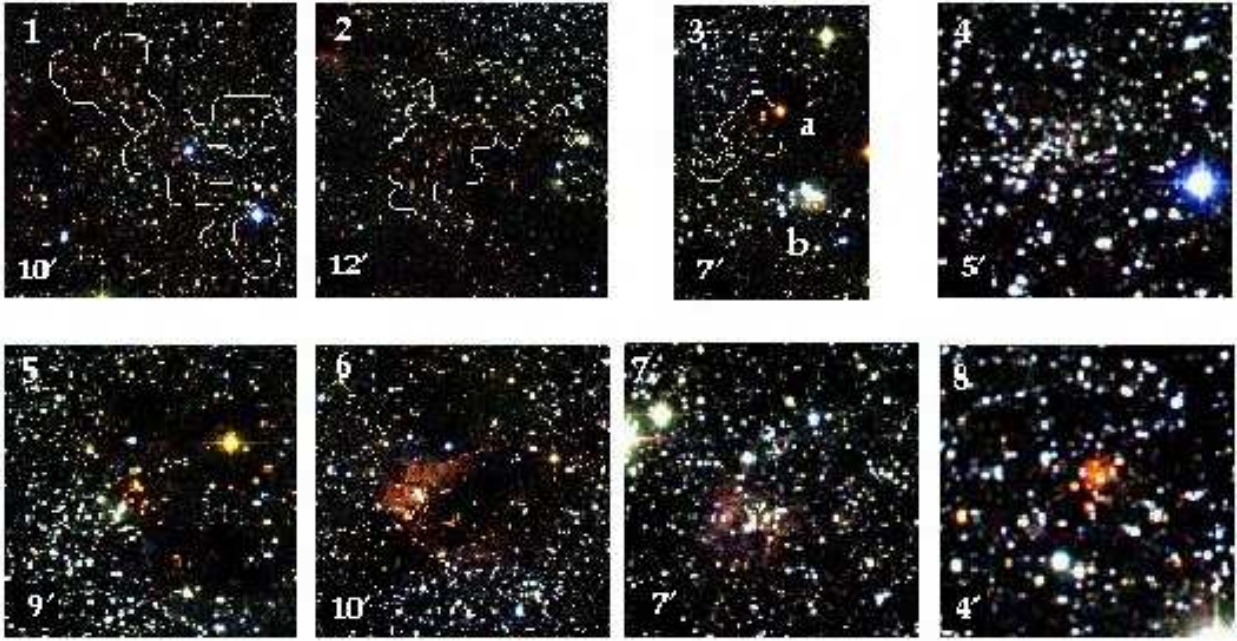}
\caption{$JHK_s$ color images of the clusters selected by comparison of the
local stellar density with the median reddening. The width of each field is
specified in the lower left corner. Contour for $A_V({\rm color})-
A_V({\rm counts})>6$~mag is overlaid when the clustering is not obvious.}
\label{clust}
\end{figure}

The technique described in section~\ref{yso} allows one to identify star
clusters by combining the color and the star count extinction maps. Since a
cluster is an excess of stars compared to the local density, it should appear
as a deficit of extinction in the star count map compared to the color map.
To identify star clusters toward the NAN, we
select areas for which $A_V({\rm color})-A_V({\rm counts}) > 6\approx 1.7\sigma$
and that contain two adjacent pixels with
$A_V({\rm color})-A_V({\rm counts}) > 14\approx 4.0\sigma$. 
These numbers were chosen subjectively after examination of the full
resolution 2MASS images. Nine cluster identifications
are proposed in Table~\ref{tab_clust}, and Figure~\ref{clust_pos} shows their
position in the cloud. Figure~\ref{clust} shows the 2MASS color images with the
threshold contour overlaid. Among these nine clusters, only number 3b and 7
were already known \citep[respectively]{Her58,WBB83}.
\begin{deluxetable}{ccc}
\tablecaption{Coordinates Of The Star Clusters\label{tab_clust}}
\tablewidth{0pt}
\tablehead{\colhead{Id} & \colhead{R.A. (J2000)} & \colhead{Dec. (J2000)}}
\startdata
1  & 20 56 02     & 43 37 00 \\
2  & 20 57 12     & 43 48 30 \\
3a & 20 58 28     & 43 56 24 \\
3b & 20 58 19     & 43 53 37 \\
4  & 20 56 13     & 44 23 05 \\
5  & 20 50 50     & 44 25 00 \\
6  & 20 53 42     & 44 31 36 \\
7  & 20 53 35     & 44 47 20 \\
8  & 20 54 14     & 44 54 07
\enddata
\end{deluxetable}

\paragraph{Cluster 1 and 2.}
They are located on the edge of the central core of the dark cloud.
The low density clustering of these red stars suggests either that the star
formation occurs in the whole core (they are detected on the edge because the
extinction is lower and the projection effect helps the detection at these
locations) or that they are in the process of being dispersed.
An upper limit for the star density excess can be obtained by assuming that
the color map gives the {\em true} extinction and that the star counts
underestimate the extinction because of the clustering. This yields only an
upper limit because young stars are known to have an intrinsic infrared
excess. For clusters 1 and 2 we find that the contour overlaid in
Figure~\ref{clust} corresponds to an excess of $\sim 4$~arcmin$^{-2}$
and that the density reaches a maximum of $\sim 20$~arcmin$^{-2}$ in their
center. For comparison, clusters 5 and 6 have a similar value for the lowest
contour but they both reach a maximum density 4 times larger of about
80~arcmin$^{-2}$.

\paragraph{Clusters 3a and 3b.}
These two clusters are on the edge of the same globule. 3a contains faint red
stars and 3b bright blue objects. Their relative position within the cloud 
may explain these colors (3a on the front side and 3b on the back side of the
cloud). 3b was discovered by \citet{Her58}.

\paragraph{Cluster 4.}
It is located in a low extinction area and stars do not exhibit any infrared
excess. It not clear whether this cluster is associated with the dark cloud;
it might be a more distant open cluster.

\paragraph{Clusters 5 and 6.}
These are the two most obvious clusters from the star count and color comparison
with $A_V({\rm color}) - A_V({\rm star count}) > 25$~mag. They both lie on the
edge of a highly extinguished globule.
Cluster 6 is associated with an emission nebula that strongly supports the
hypothesis of association with the cloud.

\paragraph{Clusters 7 and 8.}
These are in the North America nebula (in the H{\sc ii} region), in a low
extinction area ($A_V \approx 3$~mag) and apparently not associated with any
extinction core. The clustering is obvious for both clusters but there is
no other proof of their association with the dark cloud at 580~pc. 
Cluster 7 has been observed by \citet{WBB83} who concluded it
is located in the Perseus spiral arm, at a kinematic distance of 5-7~kpc,
far behind the NAN.
Cluster 8 is very compact compared to the others and that might indicate
a larger distance. It contains 12 stars among which 3 are detected in $JHK_s$,
3 in $HK_s$ and 6 in $K_s$ only.

\section{Conclusion}
We have proposed a method to map extinction with 2MASS data using
simultaneously $K_s$ star counts and $H-K_s$ reddening.
Finding a value of the extinction significantly larger from the count map
than from the color map indicates significant foreground star contamination.
Foreground stars can be removed individually in high extinction regions and
statistically elsewhere. We stress the point that the foreground population 
dominates for high extinction and these regions would not be correctly
analyzed if foreground stars are not carefully removed.
On the other hand, a higher extinction value derived from colors than from
counts indicates the presence of a star cluster. The adaptive method increases
the angular resolution of the mapping around these objects.
Color, especially $H-K_s$, is generally a better estimator of the
extinction than star counts which have a higher statistical uncertainty.

Towards the NAN, we estimate the foreground
star density to $1900\pm 500$ stars~deg$^{-2}$ and the visual extinction is
mapped up to 35~mag. Angular resolution of the extinction map
depends on the local stellar density and is $1.2'$, $3.7'$ and $7'$ for
visual extinctions of about 0, 20 and $>30$~mag, respectively.
Nine clusters are identified, only two of which were already
known and another two of which correspond to a diffuse excess of red stars close
the center of the cloud, suggesting an {\em evolved} young population or star
forming activity in the whole core ($\sim 5$~pc diameter).

Near-infrared wavelengths are required to investigate such highly obscured
fields. With the help of the accurate 2MASS photometry,
we have obtained results that supersede the previous work on the extinction
in this area. The method developed here can be extended to other high
extinction clouds with a significant fraction of foreground stars (i.e.
located at more than 500~pc, in the galactic plane) that were not accessible
before.
The knowledge of the extinction toward the galactic plane is essential in
order to constrain stellar population model parameters, such as the diffuse
extinction and the giant star distribution, that are not well constrained
at higher galactic latitudes (see appendix). 
The generalization of studies based on the comparison of the extinction
(derived from star counts and color) with molecular lines and/or
far--infrared and submillimeter emission is now possible with the 2MASS
data releases.

\acknowledgments
This publication makes use of data products from the Two Micron All Sky Survey,
which is a joint project of the University of Massachusetts and the Infrared
Processing and Analysis Center/California Institute of Technology, funded by
the National Aeronautics and Space Administration and the National Science
Foundation.\\
\indent L. Cambr\'esy acknowledges partial support from the Lavoisier grant
of the French Ministry of Foreign Affairs.\\

\appendix
\section{Comparison with a galactic stellar distribution model}
The extinction calibration is based on the $H-K_s$ color in the reference
field. This field is assumed to be representative of the aggregate color over
the field containing the NAN. Deviations from this color are
interpreted as extinction from the dust associated with the complex.
To validate this procedure, we have compared the 2MASS star counts with
the predicted counts based on galactic stellar distribution model, which
predicts both the star counts and general interstellar medium extinction for
the region of the Milky Way that includes the NAN. 
After incorporating the derived NAN extinction map with the model star counts,
the extincted star counts should match the observed 2MASS star counts
(\S~\ref{A3}). 

\subsection{The Model}
\label{A1}
The genesis of stellar distribution model comes from the \citet{BS80a}
optical star count model.  \citet{Jar92} modified the \citeauthor{BS80a}
model using the discrete formalism of \citet{Eli78}, \citet{JAHR81} and
\citet{GJ87}, extending the model to the near-infrared (1-5~$\mu$m).
The model includes the class III evolved giants, class IV subdwarfs, class V
main sequence, brown dwarfs (types L and T), and AGB populations.  As with
the \citeauthor{BS80a} model, the stars are distributed in two primary
large-scale components: disk (exponential profile) and spheroid ($R^{1/4}$
profile). Colors, luminosity, and number density per spectral type are based
on the empirical data from \citet{Wie74}, \citet{WJK83}, \citet{Koo83},
\citet{RG82}, \citet{HB88}, \citet{Bes90}, \citet{BB88}.
The interstellar extinction is applied as a smooth exponential function of
galactic position, characterized by its own scale height and disk length.
In total there are several adjustable parameters, including
disk/spheroid scale lengths, scale heights, luminosity function and dispersion,
and extinction/reddening laws. The model parameters were tuned using deep
optical and infrared star counts (see below).

\subsection{Validation of the Model}
\begin{figure}
\plotone{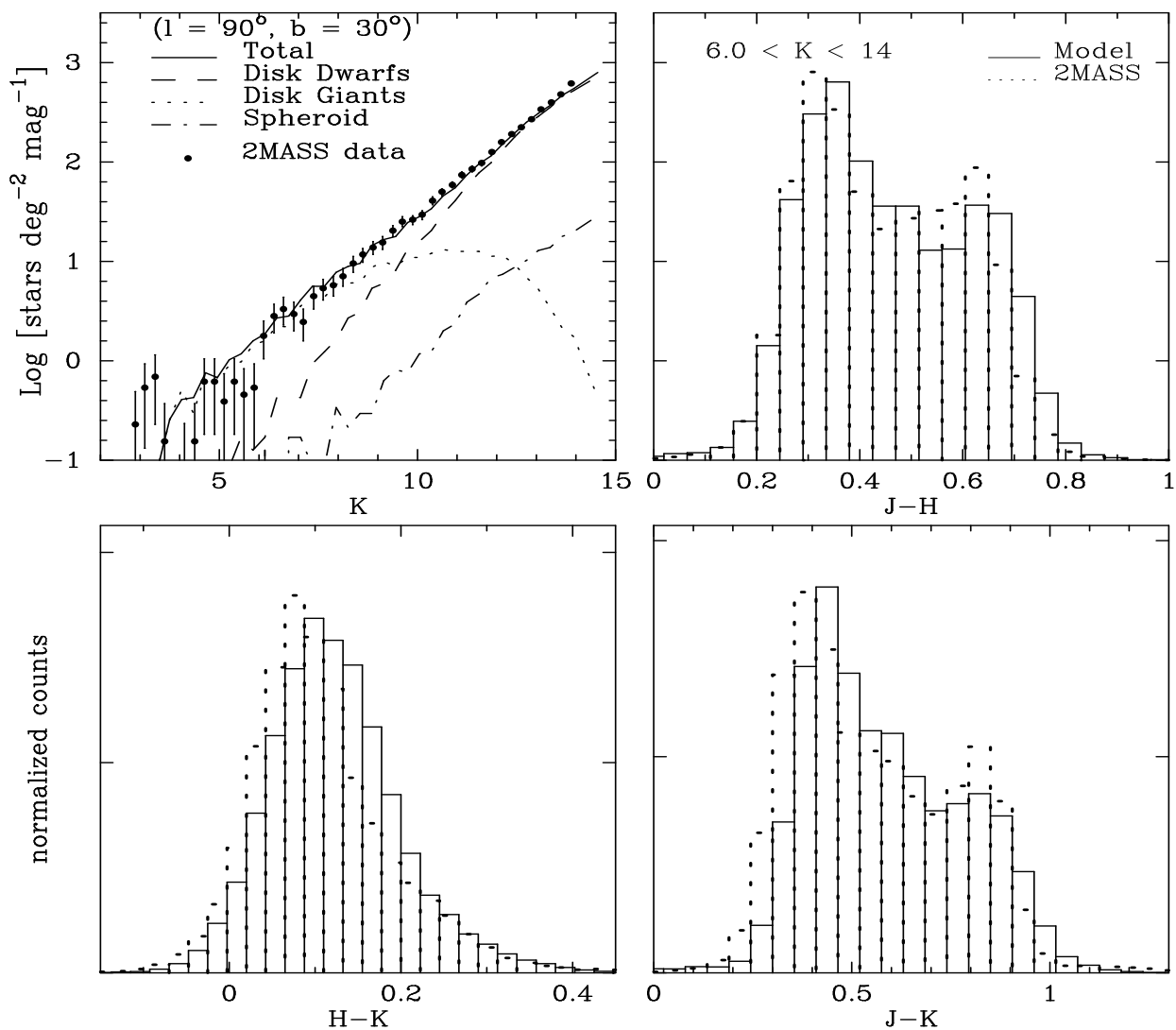}
\caption{Star counts and colors for $l = 90^\circ$, $b = 30^\circ$. A
$5^\circ \times 5^\circ$ 2MASS field, centered on galactic position
$l=90^\circ$ $b=+30^\circ$, is depicted with $K_s$-band star counts and $J-H$,
$H-K_s$ and $J-K_s$ colors. The color histograms are restricted to
$6<K_s<14$~mag. Overlaid are the model predictions.}
\label{figA1}
\end{figure}

The optical portion of the model was {\em tuned} and validated using deep CCD
observations, $V<24$ mag, $R<22$ mag, of a diverse set of fields,
encompassing both low and high stellar number density \citep{Jar92,JDH94}.
The near-infrared portion was tuned and validated using 2MASS star counts,
for $J<15.5$ and $K_s<14$ mag, for fields ranging from
$0^\circ <|b|< 90^\circ$ (Table~\ref{tableA1}). The typical field size
was 25~deg$^2$. The model performs well for $|b| > 20^\circ$. An illustrative
example of the performance for $l=90^\circ$ $b=30^\circ$ is given in
Figure~\ref{figA1}. 
At low galactic latitudes the model has small but significant differences
compared with 2MASS, probably due to differential extinction from clouds
along the line of sight (e.g., see $|b|<5^\circ$ in Table~\ref{tableA1},
corresponding to regions of the plane that are in close proximity to the
NAN). A direct comparison with the NAN is given in \S~\ref{A3}.

\begin{deluxetable}{cclcc}
\tablecaption{2MASS Fields Used To Constrain Model Parameters\label{tableA1}}
\tablewidth{0pt}
\tablehead{
\colhead{$l (^\circ)$} & \colhead{$b (^\circ)$} & \colhead{note} & \multicolumn{2}{c}{Performances} \\
\colhead{} & \colhead{} & \colhead{} & \colhead{star counts} & \colhead{color}}
\startdata
90  & 90   & Galactic North Pole & nominal & nominal\\
90  & 50   &                     & nominal & nominal\\
90  & 30   &                     & nominal & nominal\\
90  & 20   &                     & nominal & (a)\\
270 & 20   &                     & nominal & nominal\\
90  & 15   &                     & nominal & (a)\\
90  & 10   &                     & (b)     & (a)\\
180 & 10   & Galactic Anticenter & nominal & nominal\\
270 & 10   &                     & nominal & nominal\\
90  & 5    &                     & (b)     & (a)\\
90  & 2    &                     & nominal & (a)\\
86  & -2.9 & reference field (Fig.~\ref{modelNAN}, & nominal & nominal \\
    &      & lower panels) & & \\
85  & 1    & opposite the NAN  & (c)     & (c)\\
\enddata
\tablecomments{{\em Star counts} refer to counts at $K_s=14^{\rm th}$~mag;
{\em colors} refer to the $J-H$, $H-K_s$ and $J-K_s$~mag differences.
A {\em nominal} result means that the model star counts are within 10\% of
the 2MASS star counts.
(a): slight color excess in the 2MASS $J-K_s$ with respect to the model colors.
(b): 2MASS $K_s$-band star counts in $\sim 20$\% excess with
respect to the model.
(c): star counts and colors are nominal only if an additional (ad hoc) source
of extinction is included in the model: $A_V = 1.0$ at 4~kpc.}
\end{deluxetable}

\subsection{Comparing the model with the 2MASS NAN observations}
\label{A3}
The galactic stellar distribution model adequately ($\sim$ 10-20\%) predicts
the number density and colors of stars located in the galactic plane, including
areas near the NAN (Table~\ref{tableA1}). For the
cloud itself, an additional source of extinction is required to match the model
with 2MASS measurements. The histogram of the extinction we obtained using
our extinction map allows us to redden statistically stars that are located
more than 580 pc from the Sun (the nominal NAN cloud distance). We
emphasize that this extinction component (attributed to the NAN cloud) is an
addition to the nominal interstellar extinction component that is built into
the model (\S~\ref{A1}).

\begin{figure}
\plotone{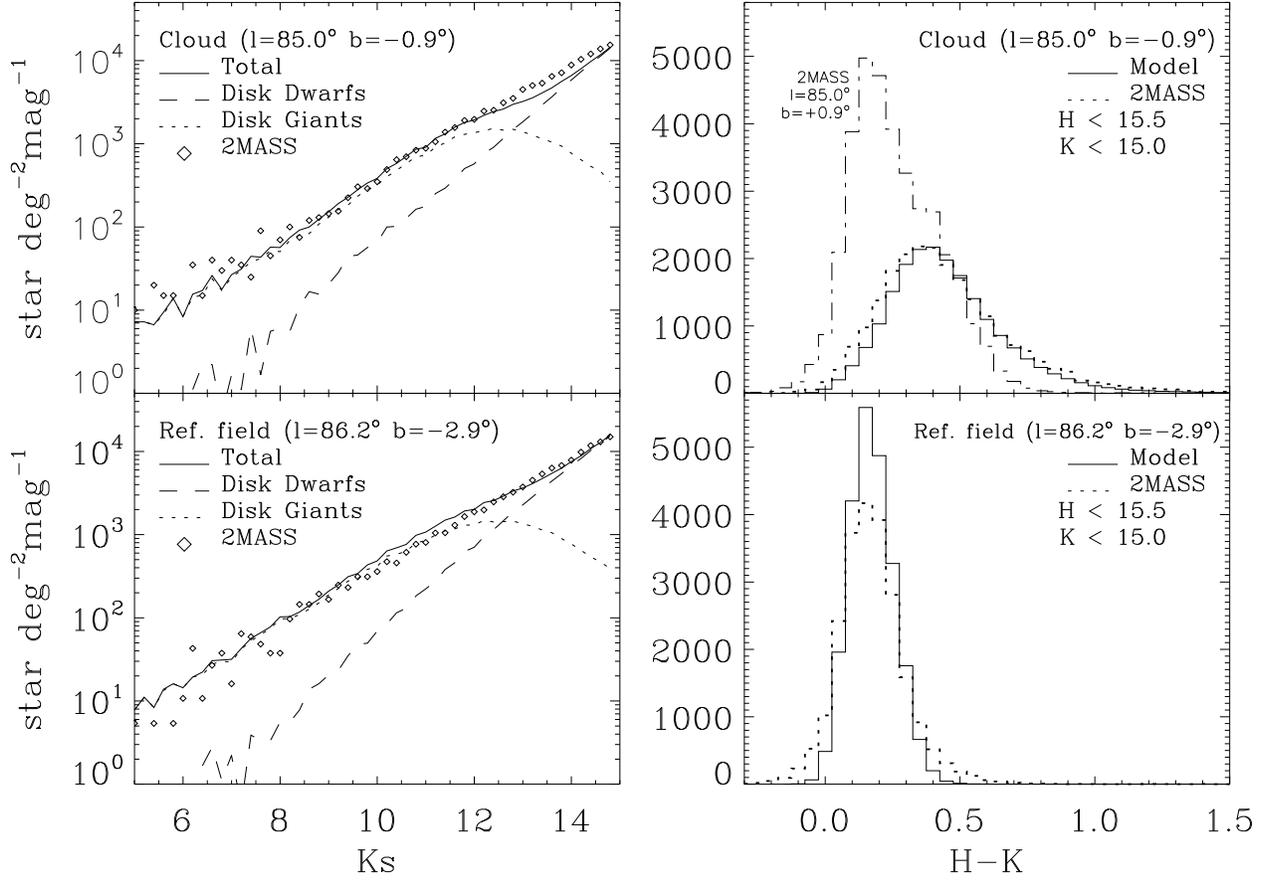}
\caption{$K_s$-band star counts and $H-K_s$ colors for the NAN
($l=85^\circ$, $b=-0.9^\circ$), upper panels, and the reference region
($l=86.2^\circ$, $b=-2.9^\circ$), lower panels. The 2MASS measurements are
depicted with diamond symbols. Overlaid (connected lines) are the model
predictions. For comparison, the upper right panel includes the predicted
$H-K_s$ colors for stars located along the line-of-sight that is directly
opposite the NAN in the galactic plane ($l=85^\circ$, $b=+0.9^\circ$; see
also Table~\ref{tableA1}). NAN model includes extinction map.}
\label{modelNAN}
\end{figure}

Figure~\ref{modelNAN} compares the 2MASS data with the extincted-corrected model
predictions.  The upper panels show the results for the cloud complex 
($l=85^\circ$, $b=-0.9^\circ$) and the lower panels show the results for the
reference field ($l=86.2^\circ$, $b=-2.9^\circ$). The lower left panel shows
differential star counts for the reference region, where the solid and dashed
lines correspond to the model and the diamonds denote the 2MASS measurements.
Note that the evolved stars dominate the bright end ($K_s < 11$~mag) and disk
dwarfs dominate the faint end. The $H-K_s$ color histograms are shown in the
lower right panel for the reference region. Both the star counts and colors
suggest that the 2MASS measurements are well fitted by the model predictions
(see also Table~\ref{tableA1}). The upper left panel shows the differential star
counts for the region along the cloud line-of-sight. The shape of the
$H-K_s$ color histograms are remarkably similar (right upper panel), again
suggesting that the model+cloud extinction adequately describes the stellar
distribution toward the NAN. For comparison we show
(upper right hand panel) the predicted stellar colors for stars located along
the line-of-sight that is directly opposite the NAN in the galactic plane
(i.e., $l= 85^\circ$, $b = +0.9^\circ$; see Table~\ref{tableA1}). Here we have
applied an ad hoc extinction, $A_V = 1$~mag at 4~kpc, to match the 2MASS star
counts along this line of sight (Table~\ref{tableA1}, note (c)). This is
consistent with the expected low extinction derived from CO emission along
this line of sight with $W({\rm CO})=11.08$~K~km~s$^{-1}$ \citep{DHT01}
which corresponds to $A_V=1.1$~mag.  
Star colors are relatively bluer and the total number of sources is nearly
double to what is seen toward the NAN.

The star count model suggests that the extinction in the reference region is
probably quite small, consistent with HI observations and visual inspection
of the POSS plates. Moreover, the predicted number of foreground stars,
$1900\pm 500$~deg$^2$, implies a cloud distance, $\sim 485$~pc, that is
comparable (within the model uncertainties, $\sim 100$~pc) to the currently
accepted distance to the NAN, $\sim 580$~pc. Given the overall high-quality
consistency between the model predictions and the 2MASS measurements, we
conclude that the reference field gives the correct zero point color in the
whole NAN complex, and the derived extinction map is a fair representative of
the dust column density attributed to the NAN.


\clearpage

\begin{thebibliography}{57}
\expandafter\ifx\csname natexlab\endcsname\relax\def\natexlab#1{#1}\fi

\bibitem[{{Alves} et~al.(2001){Alves}, {Lada}, \& {Lada}}]{ALL01}
{Alves}, J., {Lada}, C.~J., \& {Lada}, E.~A. 2001, Nature, 409, 159

\bibitem[{{Alves} et~al.(1998){Alves}, {Lada}, {Lada}, {Kenyon}, \&
  {Phelps}}]{ALL+98}
{Alves}, J., {Lada}, C.~J., {Lada}, E.~A., {Kenyon}, S.~J., \& {Phelps}, R.
  1998, ApJ, 506, 292

\bibitem[{{Andreazza} \& {Vilas-Boas}(1996)}]{AV96}
{Andreazza}, C.~M. \& {Vilas-Boas}, J. W.~S. 1996, A\&AS, 116, 21

\bibitem[{{Bahcall} \& {Soneira}(1980)}]{BS80a}
{Bahcall}, J.~N. \& {Soneira}, R.~M. 1980, ApJS, 44, 73

\bibitem[{{Bally} \& {Scoville}(1980)}]{BS80b}
{Bally}, J. \& {Scoville}, N.~Z. 1980, ApJ, 239, 121

\bibitem[{{Beichman} \& {Jarrett}(1994)}]{BJ94}
{Beichman}, C.~A. \& {Jarrett}, T. 1994, Ap\&SS, 217, 207

\bibitem[{{Bessell}(1990)}]{Bes90}
{Bessell}, M.~S. 1990, PASP, 102, 1181

\bibitem[{{Bessell} \& {Brett}(1988)}]{BB88}
{Bessell}, M.~S. \& {Brett}, J.~M. 1988, PASP, 100, 1134

\bibitem[{{Bok}(1956)}]{Bok56}
{Bok}, B.~J. 1956, AJ, 61, 309

\bibitem[{{Burton} \& {Hartmann}(1994)}]{BH94}
{Burton}, W.~B. \& {Hartmann}, D. 1994, in ASP Conf. Ser. 67: Unveiling
  Large-Scale Structures Behind the Milky Way

\bibitem[{{Cambr\'esy}(1999)}]{Cam99a}
{Cambr\'esy}, L. 1999, A\&A, 345, 965

\bibitem[{{Cambr\'esy} et~al.(1997){Cambr\'esy}, {Epchtein}, {Copet}, {de
  Batz}, {Kimeswenger}, {Le Bertre}, {Rouan}, \& {Tiph\`ene}}]{CEC+97}
{Cambr\'esy}, L., {Epchtein}, N., {Copet}, E., {de Batz}, B., {Kimeswenger},
  S., {Le Bertre}, T., {Rouan}, D., \& {Tiph\`ene}, D. 1997, A\&A, 324, L5

\bibitem[{Cardelli et~al.(1989)Cardelli, Clayton, \& Mathis}]{CCM89}
Cardelli, J.~A., Clayton, G.~C., \& Mathis, J.~S. 1989, ApJ, 345, 245

\bibitem[{{Carpenter}(2000)}]{Car00}
{Carpenter}, J.~M. 2000, AJ, 120, 3139

\bibitem[{{Cernicharo} \& {Bachiller}(1984)}]{CB84}
{Cernicharo}, J. \& {Bachiller}, R. 1984, A\&AS, 58, 327

\bibitem[{{Cutri} et~al.(2000){Cutri}, {Skrutskie}, {Van Dyk}, {Chester},
  {Evans}, {Fowler}, {Gizis}, {Howard} et~al.}]{CSV+00}
{Cutri}, R.~M., {Skrutskie}, M.~F., {Van Dyk}, S., {Chester}, T., {Evans}, T.,
  {Fowler}, J., {Gizis}, J., {Howard}, E., et~al. 2000, Explanatory Supplement
  to the 2MASS Second Incremental Data Release

\bibitem[{{Dame} et~al.(2001){Dame}, {Hartmann}, \& {Thaddeus}}]{DHT01}
{Dame}, T.~M., {Hartmann}, D., \& {Thaddeus}, P. 2001, ApJ, 547, 792

\bibitem[{{Della Prugna} et~al.(1984){Della Prugna}, {Calvet}, \&
  {Araque}}]{DCA84}
{Della Prugna}, F., {Calvet}, N., \& {Araque}, M. D.~C. 1984, Revista Mexicana
  de Astronomia y Astrofisica, 9, 31

\bibitem[{{Dickman}(1978)}]{Dic78}
{Dickman}, R.~L. 1978, AJ, 83, 363

\bibitem[{{Elias}(1978)}]{Eli78}
{Elias}, J.~H. 1978, ApJ, 223, 859

\bibitem[{{Epchtein}(1997)}]{Epc97}
{Epchtein}, N. 1997, in The Impact of Large Scale Near--Infrared Sky Surveys,
  eds. F.~{Garz\'on}, N.~{Epchtein}, A.~{Omont}, B.~{Burton}, \& P.~{Persi}
  (Tenerife: Kluwer Academic Publishers)

\bibitem[{{Feldt} \& {Wendker}(1993)}]{FW93}
{Feldt}, C. \& {Wendker}, H.~J. 1993, A\&AS, 100, 287

\bibitem[{{Fountain} et~al.(1983){Fountain}, {Gary}, \& {O'Dell}}]{FGO83}
{Fountain}, W.~F., {Gary}, G.~A., \& {O'Dell}, C.~R. 1983, ApJ, 269, 164

\bibitem[{{Frerking} et~al.(1982){Frerking}, {Langer}, \& {Wilson}}]{FLW82}
{Frerking}, M.~A., {Langer}, W.~D., \& {Wilson}, R.~W. 1982, ApJ, 262, 590

\bibitem[{{Garwood} \& {Jones}(1987)}]{GJ87}
{Garwood}, R. \& {Jones}, T. 1987, PASP, 99, 453

\bibitem[{{Goudis} \& {White}(1979)}]{GW79}
{Goudis}, C. \& {White}, N.~J. 1979, A\&A, 78, 373

\bibitem[{{Gregorio Hetem} et~al.(1988){Gregorio Hetem}, {Sanzovo}, \&
  {L\'epine}}]{GSL88}
{Gregorio Hetem}, J.~C., {Sanzovo}, G.~C., \& {L\'epine}, J. R.~D. 1988, A\&AS,
  76, 347

\bibitem[{{Hawkins} \& {Bessell}(1988)}]{HB88}
{Hawkins}, M.~R.~S. \& {Bessell}, M.~S. 1988, MNRAS, 234, 177

\bibitem[{{Herbig}(1958)}]{Her58}
{Herbig}, G.~H. 1958, ApJ, 128, 259

\bibitem[{{Jarrett}(1992)}]{Jar92}
{Jarrett}, T.~H. 1992, An optical study of the faint end of the stellar
  luminosity function, Ph.D. thesis, Massachusetts Univ., Amherst.

\bibitem[{{Jarrett} et~al.(1994){Jarrett}, {Dickman}, \& {Herbst}}]{JDH94}
{Jarrett}, T.~H., {Dickman}, R.~L., \& {Herbst}, W. 1994, ApJ, 424, 852

\bibitem[{{Jones} et~al.(1981){Jones}, {Ashley}, {Hyland}, \&
  {Ruelas-Mayorga}}]{JAHR81}
{Jones}, T.~J., {Ashley}, M., {Hyland}, A.~R., \& {Ruelas-Mayorga}, A. 1981,
  MNRAS, 197, 413

\bibitem[{{Jones} et~al.(1984){Jones}, {Hyland}, \& {Bailey}}]{JHB84}
{Jones}, T.~J., {Hyland}, A.~R., \& {Bailey}, J. 1984, ApJ, 282, 675

\bibitem[{{Koornneef}(1983)}]{Koo83}
{Koornneef}, J. 1983, A\&A, 128, 84

\bibitem[{{Kramer} et~al.(1999){Kramer}, {Alves}, {Lada}, {Lada}, {Sievers},
  {Ungerechts}, \& {Walmsley}}]{KAL+99}
{Kramer}, C., {Alves}, J., {Lada}, C.~J., {Lada}, E.~A., {Sievers}, A.,
  {Ungerechts}, H., \& {Walmsley}, C.~M. 1999, A\&A, 342, 257

\bibitem[{{Lada} et~al.(1999){Lada}, {Alves}, \& {Lada}}]{LAL99}
{Lada}, C.~J., {Alves}, J., \& {Lada}, E.~A. 1999, ApJ, 512, 250

\bibitem[{{Lada} et~al.(1994){Lada}, {Lada}, {Clemens}, \& {Bally}}]{LLCB94}
{Lada}, C.~J., {Lada}, E.~A., {Clemens}, D.~P., \& {Bally}, J. 1994, ApJ, 429,
  694

\bibitem[{{Lynds}(1962)}]{Lyn62}
{Lynds}, B.~T. 1962, ApJS, 7, 1

\bibitem[{{Maddalena} et~al.(1986){Maddalena}, {Morris}, {Moscowitz}, \&
  {Thaddeus}}]{MMMT86}
{Maddalena}, R.~J., {Morris}, M., {Moscowitz}, J., \& {Thaddeus}, P. 1986, ApJ,
  303, 375

\bibitem[{{Marcy}(1980)}]{Mar80}
{Marcy}, G.~W. 1980, AJ, 85, 230

\bibitem[{{Mattila}(1986)}]{Mat86}
{Mattila}, K. 1986, A\&A, 160, 157

\bibitem[{{Padoan} et~al.(1997{\natexlab{a}}){Padoan}, {Jones}, \&
  {Nordlund}}]{PJN97}
{Padoan}, P., {Jones}, B. J.~T., \& {Nordlund}, A.~P. 1997{\natexlab{a}}, ApJ,
  474, 730

\bibitem[{{Padoan} et~al.(1997{\natexlab{b}}){Padoan}, {Nordlund}, \&
  {Jones}}]{PNJ97}
{Padoan}, P., {Nordlund}, A., \& {Jones}, B. J.~T. 1997{\natexlab{b}}, MNRAS,
  288, 145

\bibitem[{{Reid} \& {Gilmore}(1982)}]{RG82}
{Reid}, N. \& {Gilmore}, G. 1982, MNRAS, 201, 73

\bibitem[{{Rieke} \& {Lebofsky}(1985)}]{RL85}
{Rieke}, G.~H. \& {Lebofsky}, M.~J. 1985, ApJ, 288, 618

\bibitem[{{Rossano}(1980)}]{Ros80}
{Rossano}, G.~S. 1980, AJ, 85, 1218

\bibitem[{{Savage} \& {Mathis}(1979)}]{SM79}
{Savage}, B.~D. \& {Mathis}, J.~S. 1979, ARA\&A, 17, 73

\bibitem[{{Schultheis} et~al.(1999){Schultheis}, {Ganesh}, {Simon}, {Omont},
  {Alard}, {Borsenberger}, {Copet}, {Epchtein} et~al.}]{SGS+99}
{Schultheis}, M., {Ganesh}, S., {Simon}, G., {Omont}, A., {Alard}, C.,
  {Borsenberger}, J., {Copet}, E., {Epchtein}, N., et~al. 1999, A\&A, 349, L69

\bibitem[{{Stepnik} et~al.(2001){Stepnik}, {Abergel}, {Bernard}, {Boulanger},
  {Cambr\'esy}, {Giard}, {Jones}, {Lamarre} et~al.}]{SAB+01}
{Stepnik}, B., {Abergel}, A., {Bernard}, J.-P., {Boulanger}, F., {Cambr\'esy},
  L., {Giard}, M., {Jones}, A., {Lamarre}, J.-M., et~al. 2001, A\&A, in press

\bibitem[{{Strai\v zys} et~al.(1993){Strai\v zys}, {Kazlauskas}, {Vansevi\v
  cius}, \& {{\v C}ernis}}]{SKVC93}
{Strai\v zys}, V., {Kazlauskas}, A., {Vansevi\v cius}, V., \& {{\v C}ernis}, K.
  1993, Baltic Astronomy, 2, 171

\bibitem[{{Thoraval} et~al.(1997){Thoraval}, {Boiss\'e}, \& {Duvert}}]{TBD97}
{Thoraval}, S., {Boiss\'e}, P., \& {Duvert}, G. 1997, A\&A, 319, 948

\bibitem[{{Welin}(1973)}]{Wel73}
{Welin}, G. 1973, A\&AS, 9, 183

\bibitem[{{Wendker} et~al.(1983){Wendker}, {Baars}, \& {Benz}}]{WBB83}
{Wendker}, H.~J., {Baars}, J. W.~M., \& {Benz}, D. 1983, A\&A, 124, 116

\bibitem[{{Westerhout}(1958)}]{Wes58}
{Westerhout}, G. 1958, Bull. Astron. Inst. Netherlands, 14, 215

\bibitem[{{Wielen}(1974)}]{Wie74}
{Wielen}, R. 1974, Highlights in Astronomy, 3, 395

\bibitem[{{Wielen} et~al.(1983){Wielen}, {Jahrei{\ss}}, \& {Kr{\"
  u}ger}}]{WJK83}
{Wielen}, R., {Jahrei{\ss}}, H., \& {Kr{\" u}ger}, R. 1983, in IAU Colloq. 76:
  Nearby Stars and the Stellar Luminosity Function

\bibitem[{{Wolf}(1923)}]{Wol23}
{Wolf}, M. 1923, Astron. Nachr., 219, 109

\end{thebibliography}


\end{document}